\documentclass[conference]{IEEEtran}
\IEEEoverridecommandlockouts

\usepackage{cite}
\usepackage{amsmath,amssymb,amsfonts}
\usepackage{algorithmic}
\usepackage{graphicx}
\usepackage{textcomp}
\usepackage{xcolor}
\usepackage{float}  
\usepackage{multicol}
\usepackage{multirow}
\usepackage{amsmath}
\usepackage{subcaption}

\usepackage{hyperref}
\usepackage{amssymb}        
\usepackage{pifont}         
\usepackage{fontawesome5}   
\hypersetup{
    colorlinks=true,
    linkcolor=red, 
    citecolor=blue, 
    urlcolor=magenta    
}
\def\BibTeX{{\rm B\kern-.05em{\sc i\kern-.025em b}\kern-.08em
    T\kern-.1667em\lower.7ex\hbox{E}\kern-.125emX}}
    
\begin{document}

\title{Lane-Wise Highway Anomaly Detection

}
\author{
\IEEEauthorblockN{
Mei Qiu\IEEEauthorrefmark{1},
William Lorenz Reindl
\IEEEauthorrefmark{2},
Yaobin Chen\IEEEauthorrefmark{1},
Stanley Chien\IEEEauthorrefmark{1},
Shu Hu\IEEEauthorrefmark{3}
}

\IEEEauthorblockA{Departments of \IEEEauthorrefmark{1}Electrical and Computer Engineering,
\IEEEauthorrefmark{2}Computer Science, \IEEEauthorrefmark{3}Computer and Information Technology}
Purdue University, USA \ \ \  
Email: \{qiu172, wreindl,  chen62, yschien, hu968\}@purdue.edu
}


\maketitle

\begin{abstract}
This paper proposes a scalable and interpretable framework for lane-wise highway traffic anomaly detection, leveraging multi-modal time series data extracted from surveillance cameras. Unlike traditional sensor-dependent methods, our approach uses AI-powered vision models to extract lane-specific features—including vehicle count, occupancy, and truck percentage—without relying on costly hardware or complex road modeling. We introduce a novel dataset containing 73,139 lane-wise samples, annotated with four classes of expert-validated anomalies: three traffic-related anomalies (lane blockage and recovery, foreign object intrusion, and sustained congestion) and one sensor-related anomaly (camera angle shift). Our multi-branch detection system integrates deep learning, rule-based logic, and machine learning to improve robustness and precision. Extensive experiments demonstrate that our framework outperforms state-of-the-art methods in precision, recall, and F1-score, providing a cost-effective and scalable solution for real-world intelligent transportation systems. Our dataset and code can be found here: \href{https://github.itap.purdue.edu/TASI/Lane-wise_traffic_AD.git}{\color{blue} Lane-wise-Traffic-AD}.


\end{abstract}


\section{Introduction} 
Traffic anomaly detection is a key function of intelligent transportation systems (ITS), supporting real-time monitoring and proactive management of safety risks, congestion, and infrastructure efficiency. As urban and highway networks grow more complex, timely detection of abnormal traffic behaviors is crucial for ensuring roadway safety and operational reliability~\cite{sun2025two}. In practice, however, even domain experts struggle to predefine all possible anomalies \cite{kim2023iterative}. Most anomaly detection models instead focus on learning a representation of normal traffic patterns and identifying anomalies as significant deviations from this norm \cite{arjunan2024real}. While extensive research has improved detection techniques, no studies address the classification or understanding of traffic anomaly types in specific scenarios.

To detect traffic anomalies in both highway and urban environments, vehicle flow, speed, and occupancy data are commonly utilized, with anomalies defined as deviations from typical traffic patterns under similar conditions. Datasets for this purpose are derived from both real-world traffic \cite{snineh2021detection} and simulated environments \cite{zhu2019traffic}. Data collection relies on a variety of sensors, including loop detectors, radar, LiDAR, and connected vehicle systems. However, these sensors are expensive to deploy and maintain, with limited coverage across road networks. Moreover, many approaches assume access to accurate road topology or GPS data, which limits scalability, especially in settings where only cameras are available. Compounding this challenge, existing datasets such as \cite{alexiadis2004next} primarily collect traffic data from highways but include only limited lane-level annotations, which are insufficient for studying detailed traffic behaviors or supporting models that require high spatial resolution and generalizability.

Despite advances in anomaly detection—including deep learning \cite{tran2024transformer}, statistical modeling \cite{kalair2021anomaly}, and hybrid frameworks \cite{djenouri2023hybrid}—these methods often lack robustness under varying traffic conditions. Algorithms tuned to specific traffic patterns or road structures (e.g., lane number, road width) typically perform well only in similar environments. This highlights the need for multi-branch detection frameworks that fuse complementary approaches to effectively detect anomalies under both road-structure-independent and structure-dependent conditions, particularly on highways.

\begin{figure*}[h]  
\centerline{\includegraphics[width=1.75\columnwidth]{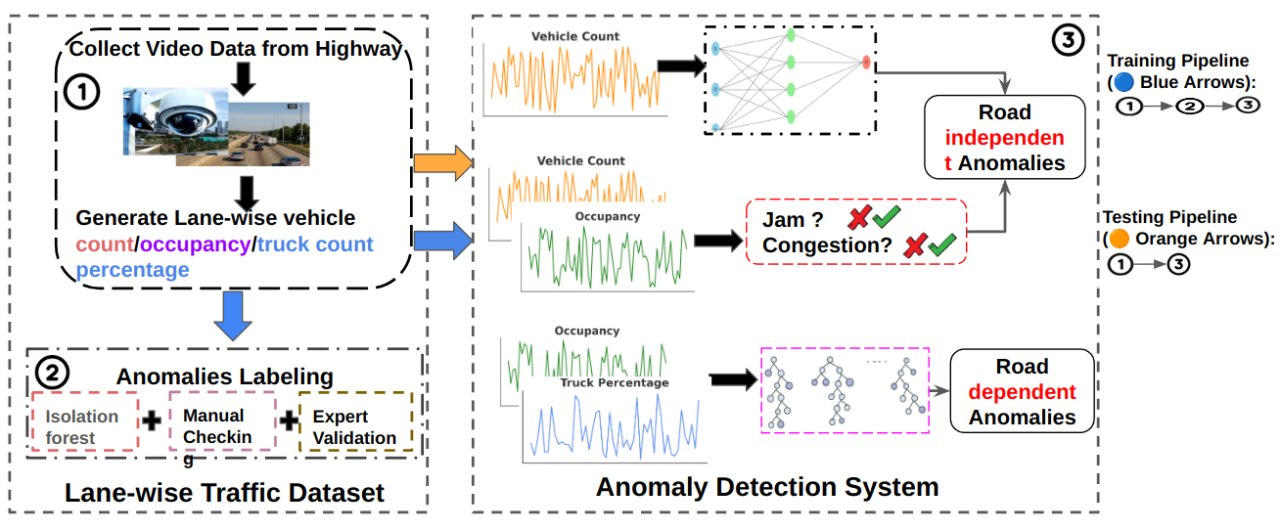}}
\vspace{-4mm}
\caption{\small Overview of the proposed lane-wise anomaly detection framework. Lane-wise traffic data (e.g., vehicle count, occupancy, and truck percentage) is extracted from highway surveillance video. Anomalies are identified through Isolation Forest, manual checking, and expert validation. The system detects road-independent anomalies using deep learning and rule-based methods, and road-dependent anomalies using machine learning models applied to occupancy and truck percentage data. The anomalies labeling is only implemented during training stage.}

\label{fig:framework}
\vspace{-4mm}
\end{figure*}
Taken together, these limitations underscore the need for a scalable, low-cost, lane-wise highway anomaly detection framework that leverages existing surveillance infrastructure to produce high-resolution, interpretable traffic signals—without reliance on expensive sensors or complex road modeling. To address this need, we propose a surveillance camera-based anomaly detection system that operates solely on roadside video feeds. The overall framework is illustrated in Fig.~\ref{fig:framework}.

Our system employs AI-powered vision models to automatically detect lane boundaries and traffic directions~\cite{qiu2024real}, and extracts lane-wise traffic features—vehicle count, occupancy, and truck percentage—from optimally learned detection regions~\cite{qiu2024intelligent}. \textbf{We focus on detecting sequential or collective traffic anomalies using these lane-wise features as input, where anomalies arise not from isolated outliers but from patterns of abnormal behavior over sequences of data points.}
To address data set limitations, we introduce a novel data set on lane-wise traffic anomalies derived from Indiana highway surveillance videos. Unlike previous data sets focused on urban intersections or aggregated data, our data set provides structured time series signals for each segment of the highway lane, supporting both lane and road anomaly detection. Through extensive data analysis and expert collaboration, we define specific types of anomalies based on the collected data. We further propose a modular, multi-stream Anomaly Detection System capable of capturing both road-independent and road-dependent anomalies.

Our contributions are summarized as follows: \begin{itemize}
\item We construct and release a novel lane-wise highway traffic dataset capturing segment-level flow dynamics from five surveillance cameras over six months, including 73,139 lane-wise samples with vehicle count, occupancy, and truck percentages. \item We propose an anomaly labeling pipeline combining machine learning, manual verification, and expert validation. Three types of traffic anomalies (lane blockage and recovery, foreign object intrusion, and sustained congestion) and one type of sensor anomaly (camera angle shift) are identified.
\item We develop a scalable, modular anomaly detection framework integrating rule-based logic, machine learning, and deep learning to improve both robustness and interpretability. \end{itemize}

\section{Related Work}
\smallskip
\noindent
\textbf{Traffic Anomaly Datasets.} Existing traffic anomaly datasets often focus on urban settings and lack relevance to highway environments. AI City Challenge~\cite{naphade2021nvidia} and GeoLife~\cite{zheng2008understanding} provide urban video and GPS data without highway-specific or anomaly-focused labeling. Highway datasets like PeMS~\cite{chen2001freeway} and NGSIM~\cite{alexiadis2004next} offer sensor and trajectory data but lack comprehensive anomaly labels and consistent lane-level detail. While XTraffic~\cite{gou2024xtraffic} and FD-AED~\cite{coursey2024ftaed} include anomaly annotations, they are limited by either spatial resolution or reliance on non-visual sensors. These limitations highlight the need for scalable, lane-wise highway datasets derived from video, enabling interpretable and generalizable anomaly detection.

\smallskip
\noindent
\textbf{Anomaly Detection on Time Series Traffic Data.} Time series anomaly detection is critical for highway traffic monitoring. Traditional statistical methods like ARIMA~\cite{cnp2024xtraffic} are simple and interpretable but struggle with nonlinear, dynamic traffic. Machine learning methods, such as Isolation forest~\cite{ liu2012isolation}, handle multivariate data and adapt better to changing patterns, yet lack deep temporal modeling. Deep generative models provide powerful and expressive tools for time series anomaly detection. Variational Autoencoders (VAE)\cite{lin2020anomaly} enhances this approach by incorporating temporal dependencies, making it more effective for sequential traffic data. Building on this, Time-VQVAE-AD~\cite{lee2024explainable} introduces vector quantization to discretize the latent space, improving both robustness and interpretability. TranAD~\cite{tuli2022tranad} takes a further step by leveraging transformer architectures, which excel at modeling long-range dependencies in multivariate time series.

\begin{figure}[h]  
\centerline{\includegraphics[width=0.8\columnwidth]{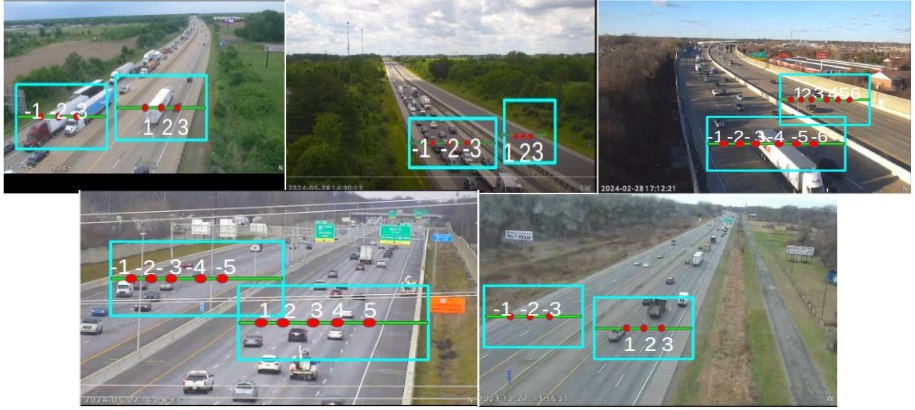}}
\caption{\small Five road views with detection regions, lane centers, and counting lines. Light blue rectangles show the learned Regions of Interest (ROIs), green lines mark optimal counting lines, and red dots indicate lane centers. White lane IDs use signs to show direction: negative for towards the camera, positive for away.}
\label{fig:cam-views}
\end{figure}

\begin{figure*}[h]  
\centerline{\includegraphics[width=2.0\columnwidth]{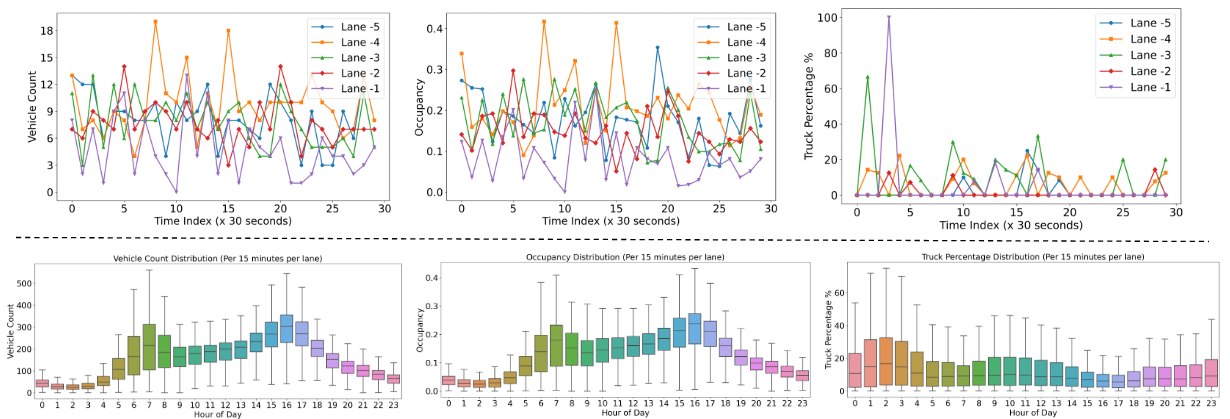}}
\caption{\small The top row shows lane-wise vehicle count, occupancy, and truck percentage from a single 15-minute video. The bottom row presents the distribution of these features across all data collected from one camera over 24 hours, aggregated across all data collecting days.}
\label{fig:cam-data}
\end{figure*}

\begin{table}[]
\caption{Summary of Traffic Datasets. (\textit{Note:} \checkmark~indicates anomaly-focused datasets, 
\faExclamationTriangle~indicates partially relevant datasets, 
and \ding{55}~indicates datasets not focused on anomaly detection.}
\label{tab:datasets}
\begin{tabular}{|p{1.0cm}|p{2.0cm}|p{1.0cm}|p{0.7cm}|p{1.8cm}|}
\hline
\textbf{Dataset} & \textbf{\begin{tabular}[c]{@{}l@{}}Sensor Type/ \\Road/\\ Anomaly-specific?\end{tabular}} & \textbf{\begin{tabular}[c]{@{}l@{}}Data\\ Size\end{tabular}} & \textbf{\begin{tabular}[c]{@{}l@{}}Lane-\\level\\Info.\end{tabular}} & \textbf{\begin{tabular}[c]{@{}l@{}}Data\\ Type\\Included\end{tabular}} \\ \hline

\begin{tabular}[c]{@{}l@{}}AI City \\ Challenge\\ \cite{naphade2021nvidia}\end{tabular} 
& \begin{tabular}[c]{@{}l@{}}Cameras (Urban)\\ Urban/ \ding{55}\end{tabular} 
& $\sim$ITB videos & No 
& \begin{tabular}[c]{@{}l@{}}Vehicle counts,\\ trajectories,\\ event\\ classification\end{tabular} \\ \hline

\begin{tabular}[c]{@{}l@{}}GeoLife\\ \cite{zheng2008understanding}\end{tabular} 
& \begin{tabular}[c]{@{}l@{}}GPS Trajectories\\ Highways/ \ding{55}\end{tabular} 
& 17,621 trajectories & No 
& \begin{tabular}[c]{@{}l@{}}User mobility\\ (location, time)\end{tabular} \\ \hline

\begin{tabular}[c]{@{}l@{}}PeMS\\ \cite{chen2001freeway}\end{tabular} 
& \begin{tabular}[c]{@{}l@{}}Loop Detectors\\ Highways/ \ding{55}\end{tabular} 
& Historical flow data & No 
& \begin{tabular}[c]{@{}l@{}}Vehicle flow,\\ occupancy, speed\end{tabular} \\ \hline

\begin{tabular}[c]{@{}l@{}}NGSIM\\ \cite{alexiadis2004next}\end{tabular} 
& \begin{tabular}[c]{@{}l@{}}Trajectory \\(video-based)/\\ Interchanges \faExclamationTriangle\end{tabular} 
& $\sim$45 min/site & Partial 
& \begin{tabular}[c]{@{}l@{}}Vehicle position,\\ speed,\\ acceleration\end{tabular} \\ \hline

\begin{tabular}[c]{@{}l@{}}XTraffic\\ \cite{gou2024xtraffic}\end{tabular} 
& \begin{tabular}[c]{@{}l@{}}Traffic Reports \\+ Cameras\\ Highways/ \checkmark\end{tabular} 
& 800K incidents & No 
& \begin{tabular}[c]{@{}l@{}}Incident types\\ (accidents,\\ congestion)\end{tabular} \\ \hline

\begin{tabular}[c]{@{}l@{}}FD-AED\\ \cite{coursey2024ftaed}\end{tabular} 
& \begin{tabular}[c]{@{}l@{}}Radar Sensors\\ Highways/ \checkmark\end{tabular} 
& $\sim$15,000 samples & Yes 
& \begin{tabular}[c]{@{}l@{}}Vehicle flow,\\ speed, \\lane occupancy\end{tabular} \\ \hline

\textbf{Ours} 
& \begin{tabular}[c]{@{}l@{}}Cameras \\(Highway video)\\ Highways/ \checkmark\end{tabular} 
& 73,139 samples & Yes 
& \begin{tabular}[c]{@{}l@{}}Vehicle flow,\\ occupancy, truck\\ percentage,\\ \textbf{anomaly types},\\ \textbf{hour labels}\end{tabular} \\ \hline

\end{tabular}
\end{table}

\begin{figure}[htbp]
    \centering

    \begin{subfigure}{0.48\linewidth}
        \centering
        \includegraphics[width=\linewidth]{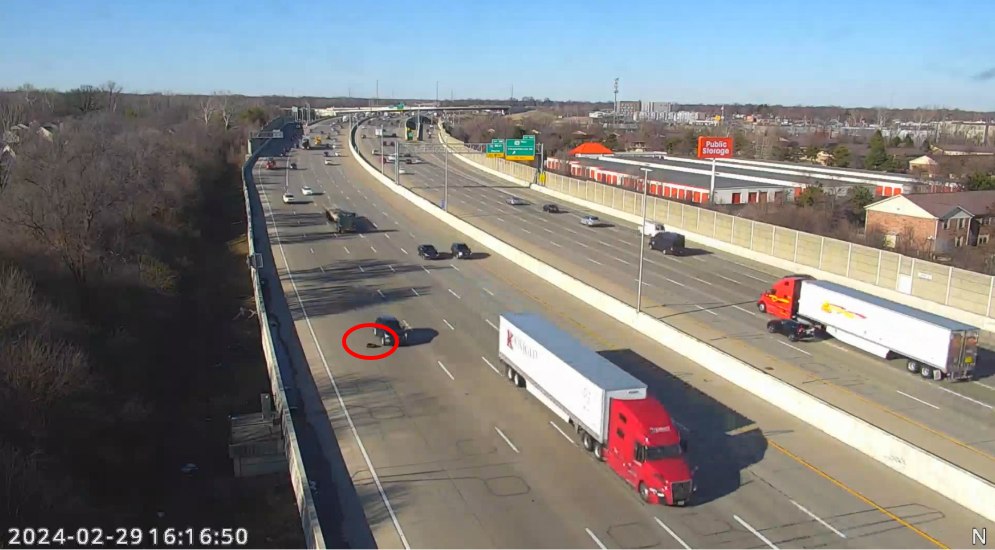}
        \caption{Foreign object (a tire) on the highway for 7 minutes (marked with a red circle)}
        \label{fig:sub1}
    \end{subfigure}
    \hfill
    \begin{subfigure}{0.48\linewidth}
        \centering
        \includegraphics[width=\linewidth]{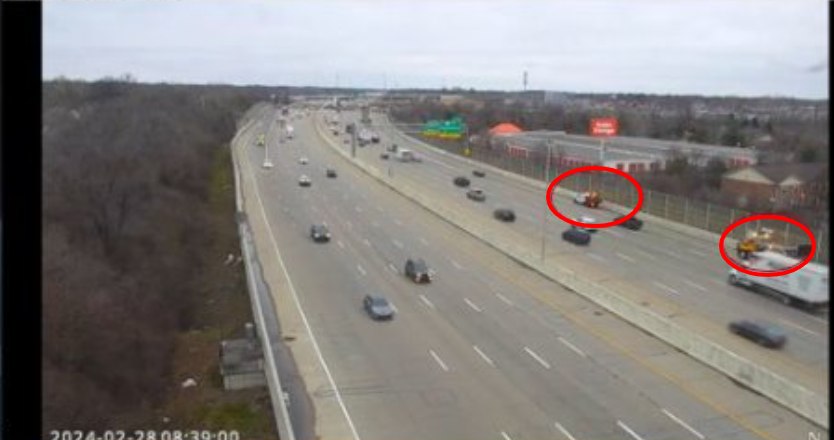}
        \caption{Truck malfunction and tow truck intervention affecting multiple lanes (marked with red circles).}
        \label{fig:sub2}
    \end{subfigure}

    \vspace{1em} 

    \begin{subfigure}{0.48\linewidth}
        \centering
        \includegraphics[width=\linewidth]{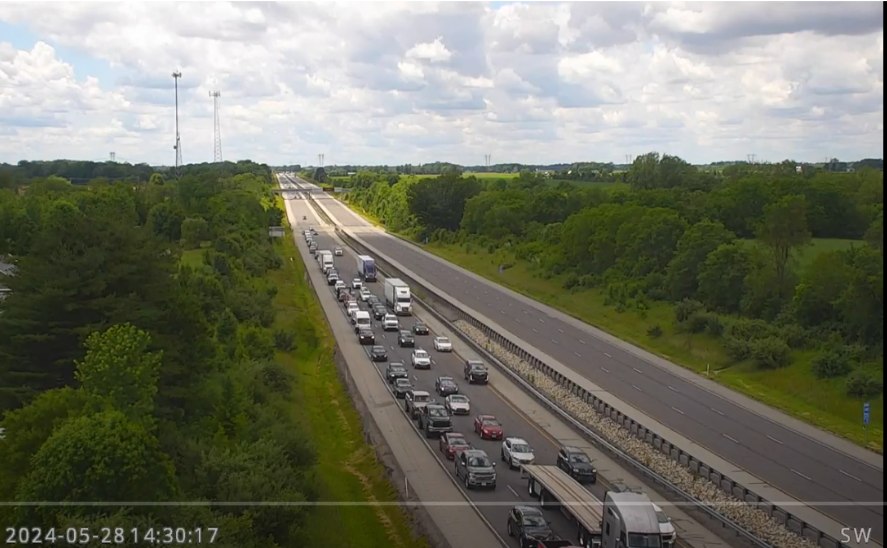}
        \caption{Vehicles only on the left roadway with congestion.}
        \label{fig:sub3}
    \end{subfigure}
    \hfill
    \begin{subfigure}{0.48\linewidth}
        \centering
        \includegraphics[width=\linewidth]{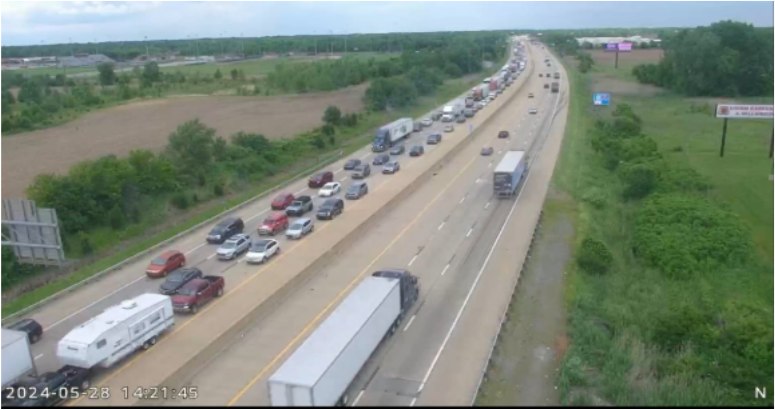}
        \caption{Traffic jam and congestion in the left lanes.}
        \label{fig:sub4}
    \end{subfigure}

    \vspace{1em}

    \begin{subfigure}{0.48\linewidth}
        \centering
        \includegraphics[width=\linewidth]{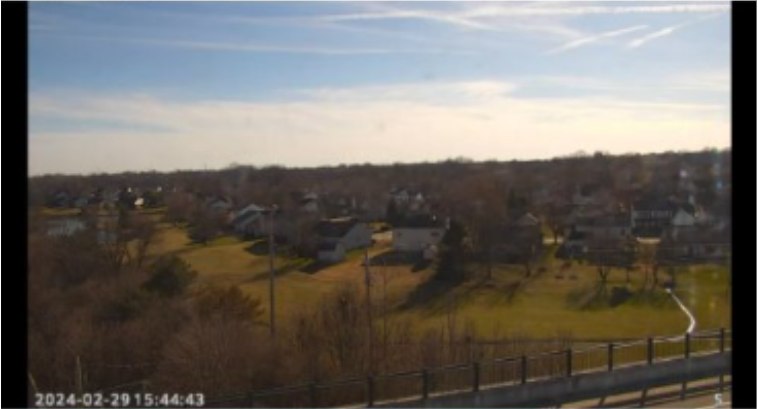}
        \caption{Camera angle shift.}
        \label{fig:sub5}
    \end{subfigure}
    \hfill
    \begin{subfigure}{0.48\linewidth}
        \centering
        \includegraphics[width=\linewidth]{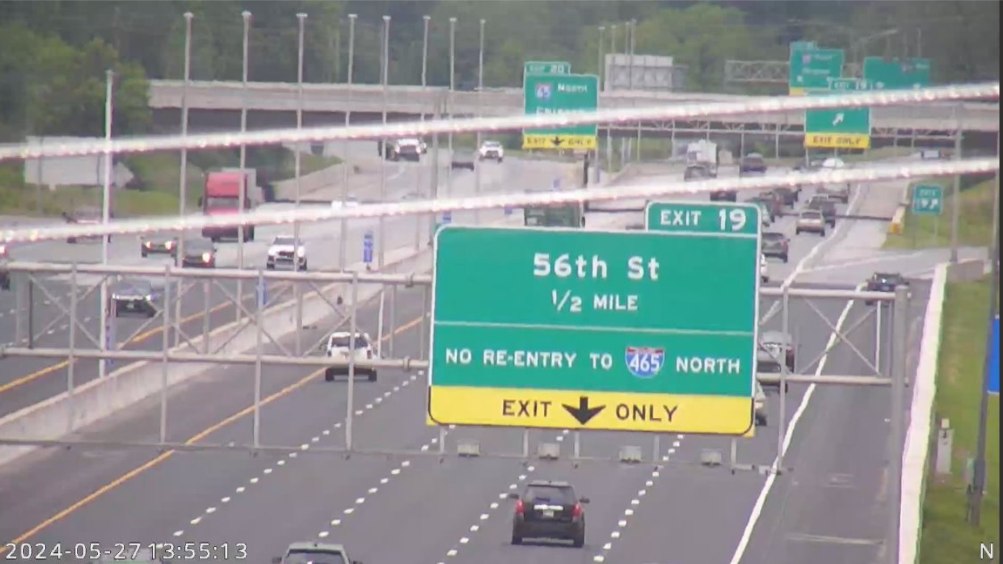}
        \caption{Camera zoom in too much.}
        \label{fig:sub6}
    \end{subfigure}

    \caption{Examples of observed traffic anomalies and sensor anomalies.}
    \label{fig:anomaly_Case}
    \vspace{-6mm}
\end{figure}

 \begin{figure*}[h]  
\centerline{\includegraphics[width=1.8\columnwidth]{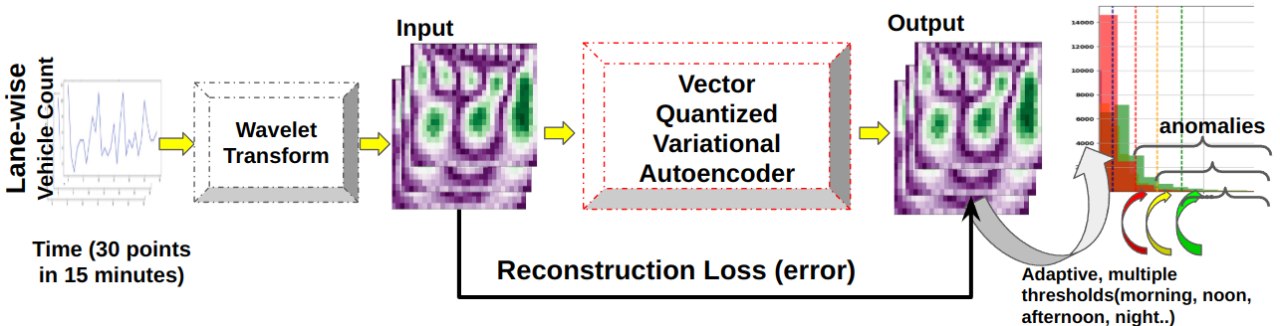}}
\vspace{-4mm}
\caption{\small Deep Learning-based Anomaly Detection method.}
\label{fig:dl_ad}
\vspace{-6mm}
\end{figure*}

\section{Dataset} 
We proposed a lane-wise traffic anomaly data collection and ground truth labeling pipeline, as shown in Fig. \ref{fig:framework}.

 

 \smallskip
\noindent
\textbf{Data Collection.} We collected a total of 8,746 videos, each 15 minutes long, from five fixed cameras installed along highways in Indiana over a six-month period. \textbf{The data collection is approved by Indiana Department of Transportation (INDOT), so that there is no ethical issue.}  Using our developed framework \cite{qiu2024intelligent}, we learned two optimal regions along with the corresponding lane centers within each region for vehicle counting. Vehicle detection was performed using YOLOv5x-CBAM \cite{qiu2022attention}, and Deep SORT \cite{wojke2017simple} was used for vehicle tracking. Only ``car" and ``truck" are detected and tracked. Fig. \ref{fig:cam-views} shows these 5 road views with the learned optimal regions, lane centers and lane identities.
Assume there are $L$ lanes learned in our observation regions (light blue rectangular in Fig. \ref{fig:cam-views}), each indexed by $i \in \{1, 2, \dots, L\}$ or $i \in \{-1, -2, \dots, -L\}$ based on traffic directions. For each lane $i$:
\begin{itemize}
    \item $c_i$: total number of vehicles detected in lane $i$ during each observation period $T$ ($T$ = 30 seconds).
    \item $t_i$: total number of trucks detected in lane $i$ during the same period.
    \item $F_{r_i}$: vehicle flow rate of lane $i$, defined as: $F_{r_i} = \frac{c_i \cdot 3600}{T}$
    (vehicles per hour).
    \item $\text{TruckPerc}_i$: truck percentage in lane $i$, computed as: $\text{TruckPerc}_i = \frac{t_i}{c_i} \times 100\%$. It is introduced to quantify the proportion of trucks relative to the total vehicle count in that lane. Incorporating this information provides richer context for distinguishing between typical fluctuations and anomalies driven by heavy vehicle influence.
    \item $\text{Occ}_{p_i}$: The occupancy contribution of vehicle $i$ is given by $\frac{h_i}{H_{\text{ROI}}}$,
\end{itemize}
where we estimate the \textbf{average lane occupancy} during the data collection interval based on the size of the region of interest (ROI) and the heights of detected vehicles passing through this ROI. Each detected vehicle contributes to occupancy proportionally to its height relative to the ROI height. Let $N$ be the number of vehicles detected during the interval $T$, $M$ be the number of frames processed within $T$, $h_i$ be the height of the bounding box for the $i$-th detected vehicle within the ROI, and $H_{\text{ROI}}$ be the height of the ROI (i.e., the vertical size of the blue rectangle shown in Fig.~\ref{fig:cam-views}). The average occupancy $\bar{O}_T$ over the interval $T$ is defined as the mean of these contributions per frame: $ \bar{O}_T = \frac{1}{M} \sum_{i=1}^{N}Occ_{p_i}.$

Throughout the remainder of this paper, occupancy refers specifically to the average lane occupancy. The upper row of Fig. \ref{fig:cam-data} shows lane-wise vehicle count, occupancy, and truck percentage from a single 15-minute video. The bottom row presents the distribution of these features across all data collected from one camera over 24 hours, aggregated across all data collecting days.

The comparison in Table~\ref{tab:datasets} highlights key limitations in existing traffic datasets for anomaly detection. Most prior datasets, such as AI City Challenge, GeoLife, and PeMS, lack anomaly-specific labels (\ding{55}) and do not offer lane-level information, limiting their utility for fine-grained behavioral analysis. NGSIM provides partial lane-level detail but remains only partially relevant to anomaly detection (\faExclamationTriangle). While XTraffic and FD-AED are anomaly-focused (\checkmark), they either lack lane resolution or rely on non-vision sensors. In contrast, our dataset is both anomaly-focused and lane-level, uniquely combining vision-based vehicle counts, occupancy, truck percentage, and expert-labeled anomaly types with temporal context---making it better suited for real-world, scalable traffic anomaly detection.

 \smallskip
\noindent
\textbf{Anomalies Labeling.} Due to the lack of a universally accepted definition of traffic anomalies, we adopt a practical approach rooted in traffic management principles to construct a verified anomaly dataset. We first apply the Isolation Forest algorithm~\cite{liu2012isolation} to identify a broad set of outlier cases from lane-wise vehicle count sequences. Each sequence $\mathbf{x}_i = [c_{i,1}, c_{i,2}, \ldots, c_{i,n}]$ represents 30 vehicle counts collected over a 15-minute video, with each $c_{i,j}$ corresponding to a 30-second interval. To ensure broad coverage of potential anomalies, we use a relatively high contamination rate (0.3). Next, normal cases are manually removed, and the remaining samples are reviewed and validated by traffic engineers. The verified anomalies are then used to train the ML-based anomaly detection model. The model assigns an anomaly score $s_i$ to each sequence: $s_i = \mathcal{F}(\mathbf{x}_i), \label{eq:iso1}$ with the following decision rule:
\begin{equation}
\text{Anomaly}(\mathbf{x}_i) =
\begin{cases}
1, & \text{if } s_i < 0, \\
0, & \text{otherwise}.
\end{cases}
\label{eq:iso2}
\end{equation}


Sequences flagged as anomalous were manually reviewed to eliminate false positives. Following this refinement, we consulted with traffic experts for validation. Based on their input, we identified and categorized \textit{three primary types of traffic anomalies and one type sensor anomaly}: (1) \textbf{Lane blockage and recovery}, typically caused by vehicle malfunctions and tow truck intervention; (2) \textbf{Foreign object intrusion}, such as a tire entering the roadway and disrupting specific lanes; (3) \textbf{Sustained congestion}, often due to truck accumulation and prolonged slow-moving traffic; and (4) sensor anomaly: \textbf{Camera angle changes or shifts.} 

From the full video dataset, 46 anomalous clips were retained after expert validation. This yielded a final dataset of 73,139 normal lane-wise samples and 341 labeled anomaly samples. Fig.~\ref{fig:anomaly_Case} illustrates representative examples of these real-world anomaly cases.
 \begin{figure}[hb]  
\centerline{\includegraphics[width=0.8\columnwidth]{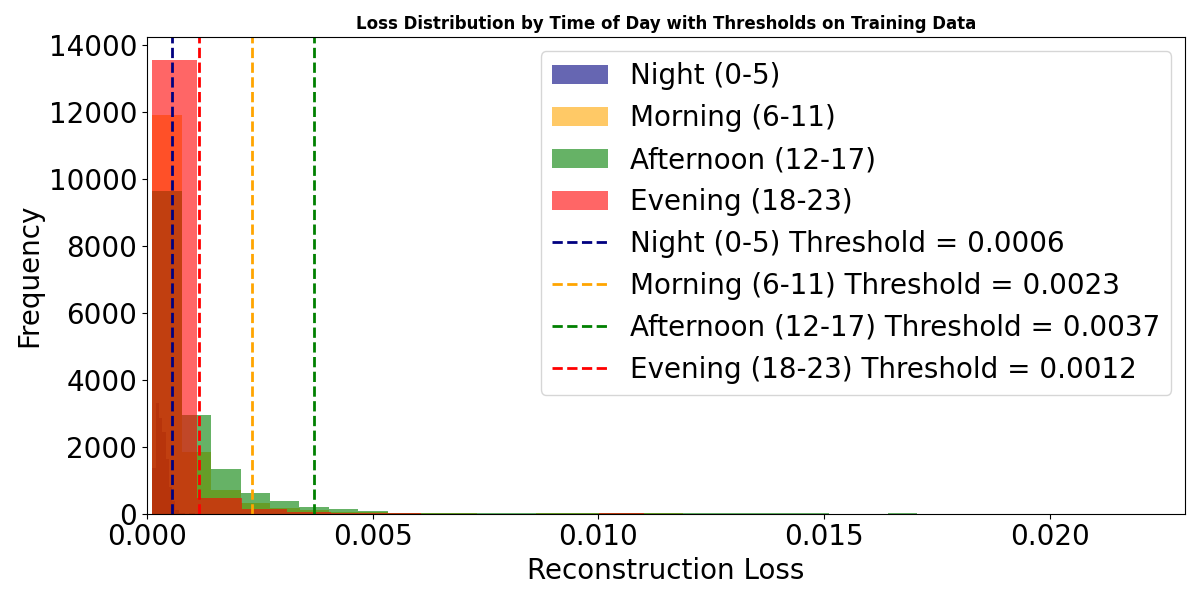}}
\caption{\small Different thresholds are computed by applying the 95th percentile to the reconstruction loss distribution within each time-of-day group (e.g., night, morning, afternoon, evening).}
\label{fig_thresholds}
\end{figure}
\section{Proposed Method}
The whole framework for the proposed fusion method is shown in Fig. \ref{fig:framework}.
\subsection{Deep Learning (DL)-based Anomaly Detection}
Fig.~\ref{fig:dl_ad} shows our deep learning-based anomaly detection framework, which identifies traffic pattern deviations across frequency bands. Lane-wise vehicle counts over 15-minute windows are converted into Continuous Wavelet Transform (CWT) spectrograms, then processed by a Vector Quantized Variational Autoencoder (VQ-VAE  \cite{van2017neural}). Anomalies are detected via reconstruction errors, using adaptive thresholds based on time-of-day groups (e.g., morning, noon). We assume minimal information loss from the CWT, that anomalies yield higher reconstruction errors, and that anomalies are sparse in training data.

\smallskip \noindent \textbf{Wavelet Transform on Traffic Data.} CWT analyzes localized variations in traffic signals across multiple frequencies. Applied to lane-specific counts, it captures both temporal and spatial patterns. Scaling adjusts frequency sensitivity—stretching reveals long-term trends like congestion, while shrinking detects abrupt events such as lane blockages—making CWT effective for detecting diverse traffic anomalies.

Mathematically, the CWT of a signal \( x(t) \) is defined as: $\text{CWT}_x(a, b) = \frac{1}{\sqrt{a}} \int_{-\infty}^{\infty} x(t) \, \psi^*\left( \frac{t - b}{a} \right) dt$,
where \( a \) is the \textit{scale parameter} (inversely proportional to frequency), \( b \) is the \textit{translation parameter} (shifting the wavelet in time), and \( \psi(t) \) is the \textit{mother wavelet}. The asterisk \( ^* \) denotes complex conjugation.

To allow different wavelet spectrograms to be directly compared in scale-independent visualizations,
we define the normalized Continuous Wavelet Transform (CWT) as follows:

\begin{equation*}
\widetilde{\text{CWT}}_x(a, b) = \frac{|\text{CWT}_x(a, b)|}{\max\limits_{a', b'} |\text{CWT}_x(a', b')|},
\end{equation*}
where $\text{CWT}_x(a, b)$ is the original complex-valued CWT at scale $a$ and time $b$, $|\cdot|$ denotes the magnitude (absolute value), the denominator $\max\limits_{a', b'} |\text{CWT}_x(a', b')|$ is the \textit{global maximum} over all scales $a'$ and translations $b'$ cross all samples, and $\widetilde{\text{CWT}}_x(a, b)$ is the \textit{normalized} wavelet coefficient, ranging from 0 to 1.

\smallskip
\noindent
\textbf{Vector Quantized Variable Autoencoder (VQ-VAE).} VQ-VAE \cite{van2017neural} is a generative model that combines the power of discrete latent representations with neural networks for efficient data reconstruction. Unlike standard VAEs that use continuous latent spaces, VQ-VAE maps inputs to a finite set of embeddings (codebook vectors). In our framework, the normalized CWT spectrogram $\widetilde{\text{CWT}}_x(a, b)$ of each lane-wise vehicle count sequence is used as the input to the VQ-VAE. Specifically, the encoder $f_{\theta}(\widetilde{\text{CWT}}_x)$ maps the normalized CWT input to a latent vector $z_e$, which is quantized to the nearest codebook vector $e_k$ from $\{e_1, e_2, \ldots, e_K\}$. The quantized latent vector $z_q = e_k$ is then passed to the decoder $g_{\phi}(z_q)$ for reconstruction. The loss function for VQ-VAE consists of three terms:
\begin{equation}
\mathcal{L} = \| \widetilde{\text{CWT}}_x - g_{\phi}(z_q) \|^2 + \| \text{sg}[z_e] - e_k \|^2 + \beta \| z_e - \text{sg}[e_k] \|^2,
\end{equation}
where $\text{sg}[\cdot]$ is the stop-gradient operator, and $\beta$ controls the commitment cost to encourage the encoder outputs to stay close to the codebook vectors. 

\smallskip
\noindent
\textbf{Anomaly Detection with Adaptive Thresholds.}
Anomalies are detected during the testing phase using dynamic thresholds derived from the reconstruction error distribution of the training data. Reconstruction error distributions from the final training epoch are grouped by time-of-day intervals: Night (0–5), Morning (6–11), Afternoon (12–17), and Evening (18–23). For each group, the 95th percentile of the corresponding error distribution is used as the anomaly detection threshold, as illustrated by the vertical dashed lines in Fig.~\ref{fig_thresholds}.

Let $\mathcal{E}_i$ denote the reconstruction error for a test sample $i$, and let $\mathcal{T}_g$ represent the threshold for the time-of-day group $g \in \{\text{Night}, \text{Morning}, \text{Afternoon}, \text{Evening}\}$. The threshold $\mathcal{T}_g$ is defined as:
\begin{equation*} \mathcal{T}g = \text{Percentile}{95}\left( { \mathcal{E}_j \mid j \in \text{Training data in group } g } \right), \end{equation*}

where $\mathcal{E}_j$ are the reconstruction errors of training samples belonging to group $g$. A test sample $i$ from group $g$ is classified as anomalous if 
\begin{equation} \mathcal{E}_i > \mathcal{T}{g}. \end{equation}



\subsection{Rule-based Anomaly Detection}
While the DL-based model effectively detects various anomalies, it may miss gradual congestion patterns that do not cause large reconstruction errors. To address this, we directly monitor flow rate $F_{r_i}$ and occupancy $\text{Occ}_{p_i}$ for each lane $i$, defining the traffic status $status_i$ as:

\begin{equation}\small
\text{status}_i =
\begin{cases}
    \text{Jam}, & \text{if } F_{r_i} < 600 \ \text{and} \ \text{Occ}_{p_i} > 0.6, \\[5pt]
    \text{Slow}, & 
    \begin{aligned}
    &\text{if } 600 < F_{r_i} < 900 \ \text{\&} \ 0.4 < \text{Occ}_{p_i} < 0.6,
    \end{aligned} \\[5pt]
    \text{Normal}, & \text{otherwise}.
\end{cases}
\end{equation}

This rule-based framework enables the detection of \textit{\textbf{jam and congestion anomalies}} by identifying lane segments exhibiting low vehicle throughput but high physical occupancy. 
\subsection{Machine Learning (ML)-based Anomaly Detection}
 \begin{figure}[h]  
\centerline{\includegraphics[width=1\columnwidth]{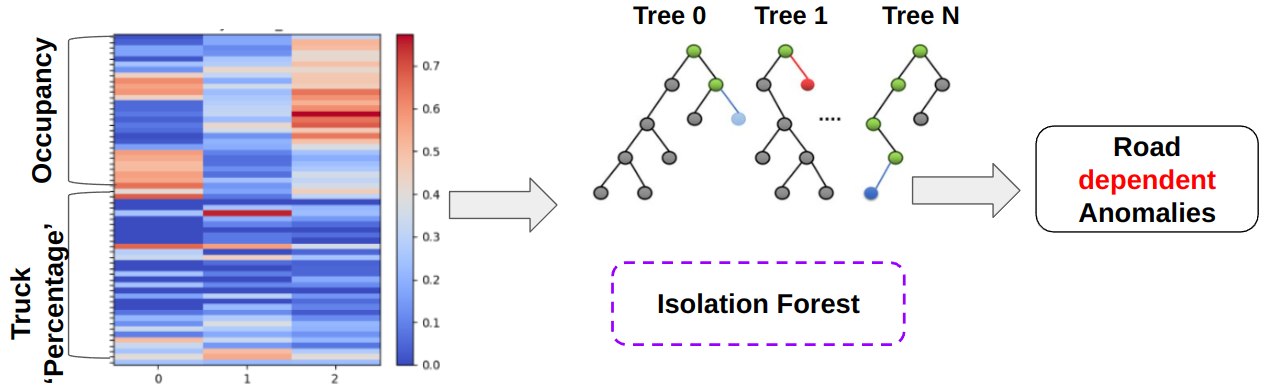}}
\caption{\small ML-based Anomaly Detection Framework.}
\label{fig:ml-ad}
\end{figure}
The ML-based module (framework shown in Fig.~\ref{fig:ml-ad}) is critical for detecting \textbf{road-dependent anomalies} that are not effectively captured by DL-based or rule-based methods. To identify complex patterns—such as truck-induced slowdowns, where heavy vehicles occupy more space and accelerate slowly, contributing to stop-and-go congestion—we incorporate both occupancy and truck percentage as input features. While occupancy alone provides partial insight, the truck percentage offers additional context to better characterize road-specific traffic disruptions.

For each video, road-dependent anomaly detection is performed at the ROI level of each traffic direction. Let the occupancy data be denoted as $\mathbf{O} \in \mathbb{R}^{N \times L}$, where $N$ is the number of time intervals and $L$ is the number of lanes. We vertically stack the occupancy data and truck percentage data $\mathbf{T} \in \mathbb{R}^{N \times L}$ to construct the feature matrix:
\begin{equation} \mathbf{X} = [\mathbf{O}; \mathbf{T}] \in \mathbb{R}^{2N \times L}, \end{equation}
where $[\cdot ; \cdot]$ denotes vertical concatenation. This feature matrix is used for road-level anomaly detection by inputting $\mathbf{X}$ into an Isolation Forest model $\mathcal{F}$ (see Eq. (\ref{eq:iso2})). The contamination rate is set to 0.1. The model outputs a binary label for each sample: a value of $-1$ indicates an anomaly, while $1$ denotes a normal case, effectively separating outliers from typical traffic patterns.

\section{Software Development}
\begin{figure}[htbp]
    \centering
    \begin{subfigure}{\linewidth}
        \centering
        \includegraphics[width=0.9\linewidth]{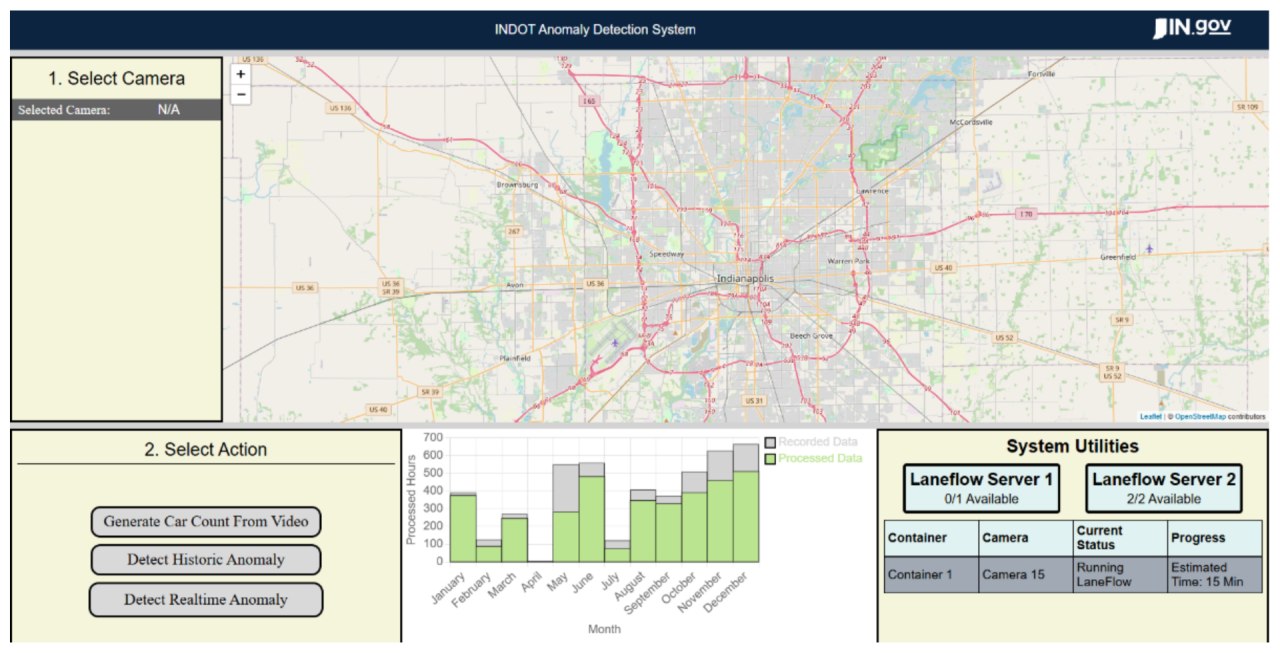}
        \caption{}
        \label{fig:a}
    \end{subfigure}
    
    \vspace{0.5em} 
    
    \begin{subfigure}{\linewidth}
        \centering
        \includegraphics[width=0.9\linewidth]{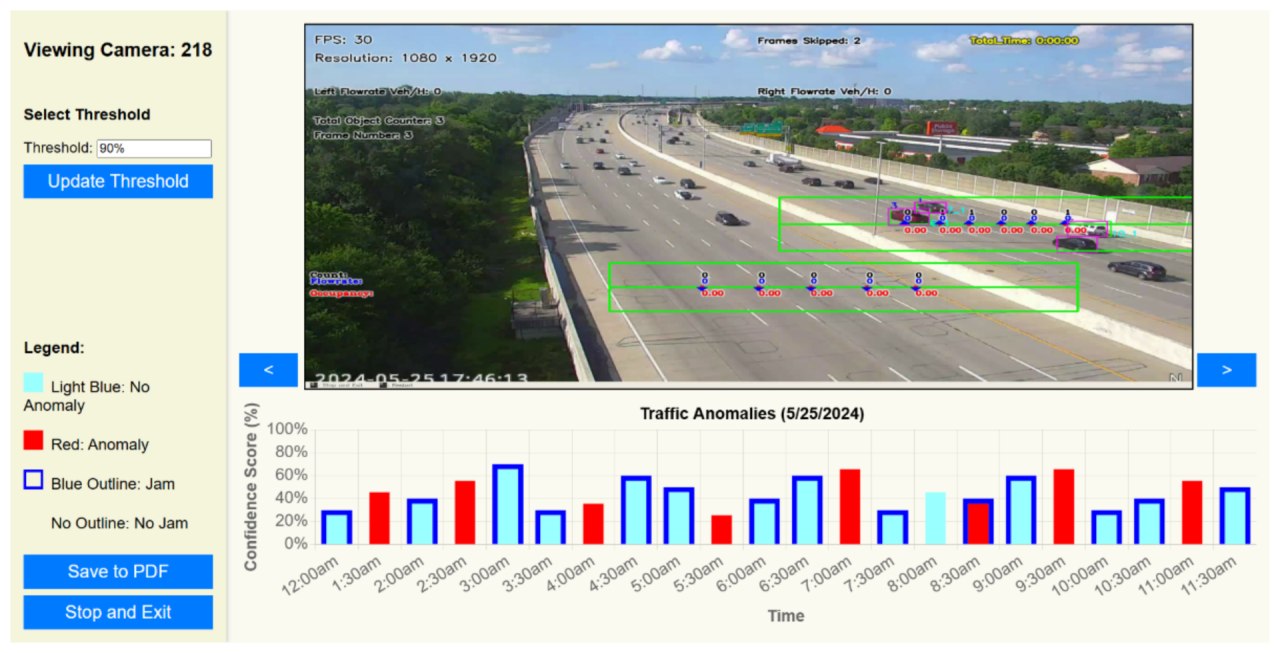}
        \caption{}
        \label{fig:b}
    \end{subfigure}
    \vspace{-5mm}
    \caption{Web-based platform for traffic monitoring and anomaly visualization. (a) Interface for accessing and processing video from over 600 highway cameras with automated lane-wise traffic feature extraction. (b) User interface for visualizing lane-level anomaly detection results.}

    \label{fig:web_gui}
    \vspace{-2mm}
\end{figure}

We developed a web-based platform (Fig.~\ref{fig:web_gui}), it enables automated video processing for vehicle detection, lane center estimation, travel direction analysis, and region-of-interest selection, allowing for accurate lane-wise extraction of vehicle counts, occupancy, and truck percentage. Additionally, we designed a graphical user interface to display lane-level anomaly detection results for each input sample. \textbf{A video demo is provided in supplementary material.}

\section{Experiments}
\subsection{Experiments Settings}
\vspace{-2mm}
\smallskip
\noindent
\textbf{Data.} 80\% of the anomaly-free lane-wise count sequences are used to train the VQ-VAE model, while the remaining 20\% serve as validation data. For testing, 341 anomalous samples from 43 videos are combined with 341 normal samples randomly selected from the validation set. This test set is used for evaluating all models (DL-based, ML-based and Rule-based).

\smallskip
\noindent



\smallskip
\noindent
\textbf{Performance Evaluation Metric.} Anomaly detection performance is commonly evaluated using several key metrics derived from the confusion matrix. \textbf{Accuracy (Acc)} measures the proportion of correctly classified instances: $\text{Accuracy} = \frac{TP + TN}{TP + TN + FP + FN}$. \textbf{Precision (Pre)} quantifies the proportion of true anomalies among all detected anomalies: $\text{Precision} = \frac{TP}{TP + FP}$. \textbf{Recall (Re)}, or sensitivity, indicates the proportion of correctly detected anomalies among all actual anomalies: $\text{Recall} = \frac{TP}{TP + FN}$. The \textbf{F1-score (F1)} provides a harmonic mean of precision and recall: $F1 = \frac{2 \times \text{Precision} \times \text{Recall}}{\text{Precision} + \text{Recall}}$. Additionally, the \textbf{False Positive Rate (FPR)} and \textbf{False Negative Rate (FNR)} are given by $\text{FPR} = \frac{FP}{FP + TN}$ and $\text{FNR} = \frac{FN}{FN + TP}$ respectively, representing the rates of incorrect anomaly detection and missed anomalies.

\smallskip
\noindent
\textbf{Implementation Details.} The VQ-VAE consists of an encoder with two Conv2d layers ($1 \times 64$ and $64 \times 32$, kernel size 4, stride 2, padding 1), each followed by BatchNorm and ReLU. A $32 \times 32$ pre-quantization Conv2d feeds into a codebook of 64 embeddings (dimension 32). Post-quantization, a $32 \times 8$ Conv2d layer prepares features for the decoder, which mirrors the encoder with two ConvTranspose2d layers ($8 \times 64$ and $64 \times 1$) and a final Tanh activation. The commitment loss weight $\beta$ is set to 0.25. The model is trained using Stochastic Gradient Descent (SGD) with a learning rate of $1 \times 10^{-3}$, weight decay of $1 \times 10^{-5}$, batch size 128, and up to 150 epochs. Early stopping is applied by monitoring validation loss to prevent overfitting. The IsolationForest model used in this work is implemented using the \texttt{IsolationForest} class from the \texttt{sklearn.ensemble} module in the scikit-learn.

\begin{table}[]
\caption{Performance comparison with other methods. \textit{Note: Best performance is highlighted in bold.}}
\label{tab:accuracy}
\centering
\begin{tabular}{|c|c|c|c|c|}
\hline
\textbf{Approach} & \textbf{Method} & \textbf{Pre} & \textbf{Re} & \textbf{F1} \\ \hline
\begin{tabular}[c]{@{}c@{}}Isolation \\ Forest\end{tabular} 
& Isolation Forest\cite{liu2012isolation} & 0.177 & 0.6047 & 0.2738 \\ \hline
\begin{tabular}[c]{@{}c@{}}Density \\ Distribution\end{tabular} 
& OWAM \cite{choudhary2023enhancing} & 0.268 & 0.2558 & 0.2558 \\ \hline
\begin{tabular}[c]{@{}c@{}}Generative \\ Modeling\end{tabular} 
& TimeVQVAE-AD \cite{lee2024explainable} & 0.45 & 0.37 & 0.4061 \\ \hline
\begin{tabular}[c]{@{}c@{}}Reconstruction \\
\& Prediction\end{tabular} 
& TranAD\cite{tuli2022tranad} & 0.32 & 0.49 & 0.3872 \\ \hline
{Reconstruction} 
& \textbf{Ours} & \textbf{0.8077} & \textbf{0.9767} & \textbf{0.8841} \\ \hline
\end{tabular}
\end{table}

\begin{table}[]
\caption{Performance with different modules. \textit{Note: Best performance is highlighted in bold.}}
\label{tab:module_comparison}
\begin{tabular}{|c|c|c|c|c|c|c|}
\hline
\textbf{Method}                                                                                        & \textbf{Acc} & \multicolumn{1}{l|}{\textbf{Pre}} & \textbf{Re} & \multicolumn{1}{l|}{\textbf{F1}} & \multicolumn{1}{l|}{\textbf{FPR}} & \multicolumn{1}{l|}{\textbf{FNR}} \\ \hline
DL-based                                                                                               & 0.5279       & 0.7368                            & 0.9767      & 0.84                             & 0.045                             & 0.023                             \\ \hline
\begin{tabular}[c]{@{}c@{}}DL-based \&\\  Rule-based\end{tabular}                                      & 0.5293       & 0.7414                            & \textbf{1.0}         & 0.851                            & 0.045                             & \textbf{0}                                 \\ \hline
\multicolumn{1}{|l|}{\begin{tabular}[c]{@{}l@{}}DL-based \& \\ Rule-based \&\\  ML-based\end{tabular}} & \textbf{0.9787}       & \textbf{0.8431}                            & \textbf{1.0}         & \textbf{0.9149}                           & \textbf{0.024}                             & \textbf{0}                                 \\ \hline
\end{tabular}
\end{table}

\vspace{-4mm}
 \begin{figure}[hb]  
\centerline{\includegraphics[width=0.8\columnwidth]{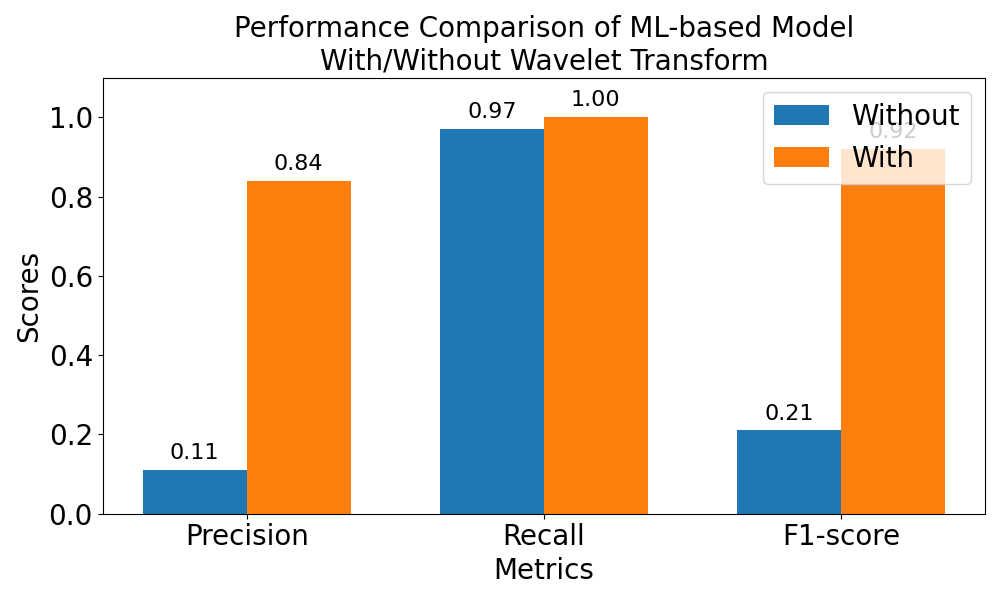}}
\vspace{-4mm}
\caption{\small Performance comparison of ML-based models with and without wavelet transform.}
\label{fig:fig_with_o_wave}
\vspace{-4mm}
\end{figure}
 \begin{figure}[h]  
\centerline{\includegraphics[width=1\columnwidth]{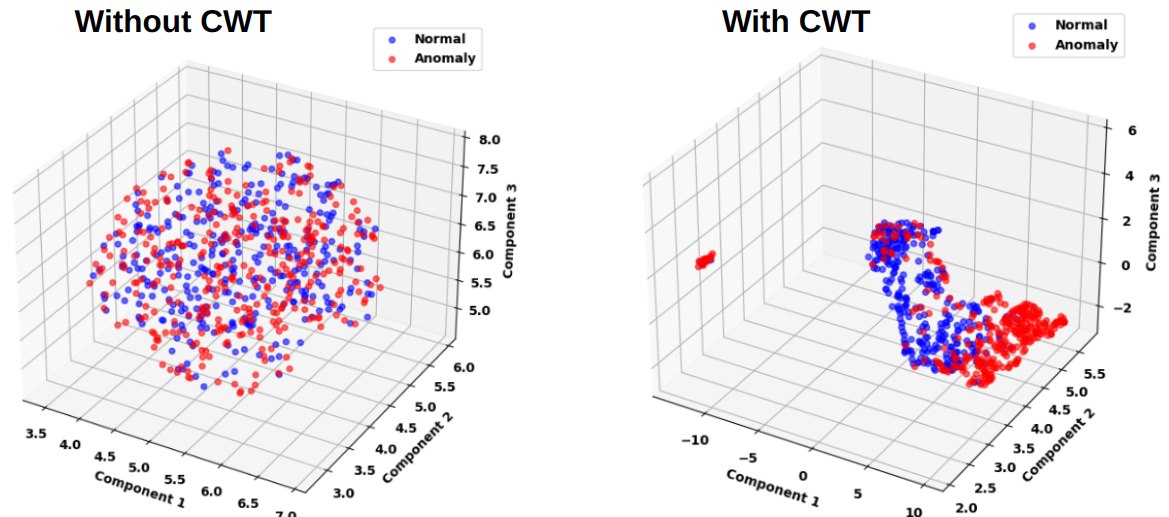}}
\caption{\small  UMAP visualization of 3D latent spaces learned by VQ-VAE. Left: Without CWT; Right: With CWT. Blue points represent normal samples, and red points represent anomalies.}
\label{fig_latent_analysis}
\end{figure}
 \begin{figure}[h]  
\centerline{\includegraphics[width=1\columnwidth]{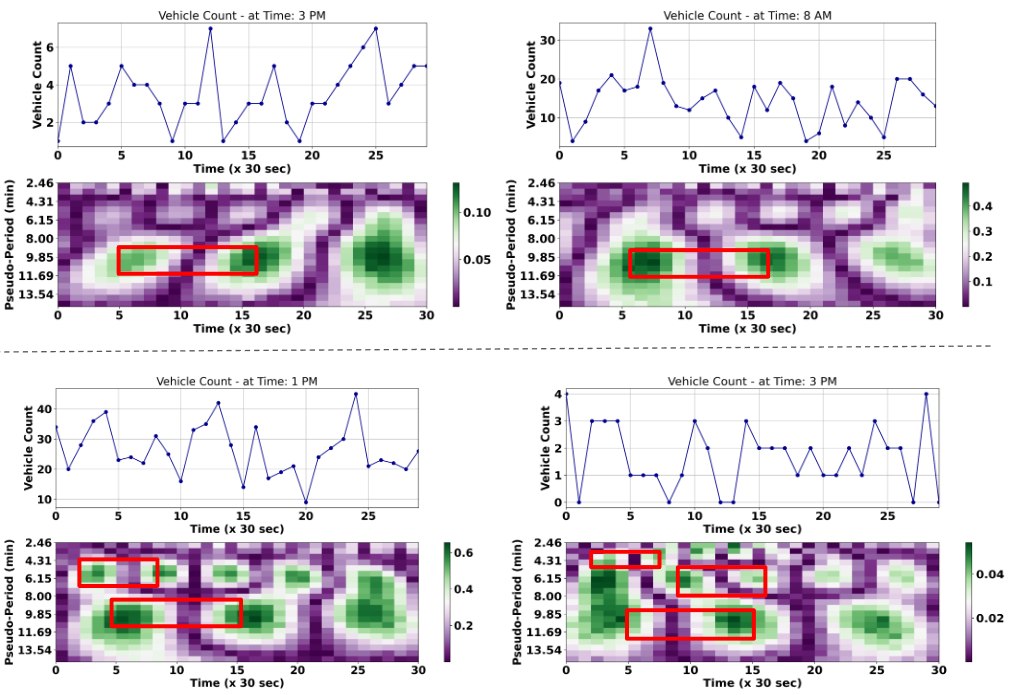}}
\caption{\small Wavelet analysis of traffic data from a single lane. The top part shows normal cases, while the bottom row illustrates anomaly cases. For each case, the upper row displays vehicle counts over a 15-minute period, and the lower panel presents the corresponding wavelet transform (normalized absolute values). Normal traffic exhibits a dominant periodic pattern around 9 minutes. In contrast, anomalous traffic shows two to three prominent periodic patterns at approximately 2, 4, and 9 minutes (marked with red rectangular).}

\label{fig:fig_cwt}
\end{figure}

\subsection{Results}
\vspace{-2mm}
\noindent
\textbf{Results compared with other methods.} Table~\ref{tab:accuracy} presents a comparison of precision, recall, and F1-score across various anomaly detection methods applied to our dataset. For each baseline method, the input consists of our lane-wise vehicle count data, with data from all lanes in each video concatenated into a single sequence. The original configurations of these methods were preserved as specified in their respective implementations. Our method outperforms all baselines (e.g., Isolation Forest\cite{liu2012isolation}, OWAM \cite{choudhary2023enhancing}, TimeVQVAE-AD \cite{lee2024explainable},  TranAD\cite{tuli2022tranad}), achieving the highest F1-score of 0.8841, indicating superior precision and recall.

\smallskip
\noindent
\textbf{Ablation Study.}  Table \ref{tab:module_comparison} highlights the effectiveness of combining detection modules. The DL-based module alone achieves an F1-score of 0.84, increasing to 0.851 when integrated with the Rule-based module. The full framework, combining DL-based, Rule-based, and ML-based modules, achieves the best results with 0.9787 accuracy and 0.9149 F1-score. Each module contributes uniquely: DL captures temporal anomalies, Rule-based detects abrupt disruptions via occupancy thresholds, and ML targets road-dependent patterns using occupancy and truck percentage. Their integration ensures robust and precise anomaly detection across varied traffic scenarios.

We also evaluate the impact of wavelet transforms in the ML-based framework. As shown in Fig.\ref{fig:fig_with_o_wave}, applying CWT significantly boosts detection performance. Fig.\ref{fig_latent_analysis} further shows that CWT-transformed inputs lead to clearer separation between normal and anomalous data in the UMAP-projected latent space, whereas models trained on raw time series inputs produce more isotropic and entangled representations, making anomaly detection more challenging.

In addition, the wavelet analysis in Fig.~\ref{fig:fig_cwt} suggests that structural differences in wavelet patterns could provide valuable cues for distinguishing normal and anomalous traffic behaviors. Normal samples typically exhibit a dominant periodic pattern (e.g., around 9 minutes), while anomalies display multiple distinct periodic components at shorter intervals (e.g., 2, 4, and 9 minutes). \textbf{This observation indicates the potential to systematically link wavelet pattern structures with anomaly characterization, offering a new direction for more interpretable anomaly detection.}

\smallskip
\noindent
\textbf{Sensitivity Analysis.} We compare the performance of DL-based anomaly detection with different threshold setting strategies in Table~\ref{tab:thresholds_ml}. Thresholds are set at the 90\%, 95\%, and 99\% percentiles of reconstruction loss from the last training epoch. In time-dependent strategies, thresholds are computed separately for different time-of-day groups (e.g., morning, afternoon, night), while the time-independent strategy uses a single threshold for all data. The 99\% time-dependent threshold yields the best performance. In contrast, the time-independent threshold strategy shows lower F1-score and higher false negative rate, demonstrating that time-aware thresholds better capture diurnal traffic variations and improve anomaly detection.


\begin{table}[]
\caption{Performance with different threshold stragies in ML-based module. \textit{Note: Best performance is highlighted in bold.}}
\label{tab:thresholds_ml}
\begin{tabular}{|c|c|c|c|c|c|c|}
\hline
\textbf{\begin{tabular}[c]{@{}c@{}}Threshold \\ Rule\end{tabular}}                                & \textbf{Acc}                & \multicolumn{1}{l|}{\textbf{Pre}} & \textbf{Re}                 & \multicolumn{1}{l|}{\textbf{F1}} & \multicolumn{1}{l|}{\textbf{FPR}} & \multicolumn{1}{l|}{\textbf{FNR}} \\ \hline
\begin{tabular}[c]{@{}c@{}}90\% (time \\ dependent)\end{tabular}                                  & 0.5117                      & 0.6143                            & \textbf{1.0}                         & 0.7611                           & 0.08                              & \textbf{0}                                 \\ \hline
\begin{tabular}[c]{@{}c@{}}95\% (time \\ dependent)\end{tabular}                                  & 0.5279                      & 0.7368                            & 0.9767                      & 0.84                             & 0.045                             & 0.023                             \\ \hline
\multicolumn{1}{|l|}{\begin{tabular}[c]{@{}l@{}}99\% (time \\ dependent)\end{tabular}}            & 0.5425                      & \textbf{0.9512}                            & 0.907                       & \textbf{0.9286}                          & \textbf{0.006}                             & 0.093                             \\ \hline
\multicolumn{1}{|l|}{\begin{tabular}[c]{@{}l@{}}99\% (time \\ independent)\end{tabular}} & \multicolumn{1}{l|}{\textbf{0.9202}} & \multicolumn{1}{l|}{0.6226}       & \multicolumn{1}{l|}{0.7674} & \multicolumn{1}{l|}{0.6872}      & \multicolumn{1}{l|}{0.06}         & \multicolumn{1}{l|}{0.23}         \\ \hline
\end{tabular}
\end{table}


\section{Discussion}
\vspace{-2mm}
\noindent
\textbf{Limitations.} This study is constrained by a geographically limited dataset, which may affect the model’s generalizability to diverse traffic environments. While a universally accepted definition of traffic anomalies is lacking, new cases can be continuously collected and incorporated into model training to improve coverage. Additionally, the current system does not include active learning, continual updates, or structured expert feedback mechanisms, which are important for adapting to evolving traffic dynamics—offering promising directions for future enhancement.

\noindent
\textbf{Future Work.} To improve robustness and scalability, future work will expand the dataset across diverse locations and road types. We plan to develop a more principled anomaly definition using hybrid labeling strategies that combine domain knowledge, behavioral clustering, and statistical thresholds. A frequency band filtering approach will be introduced to emphasize temporal patterns relevant to anomalies. Additionally, human-in-the-loop components—such as expert review, active learning, and continual model updates—will support adaptive and interpretable anomaly detection.

\section{Conclusion}
\vspace{-2mm}
In this paper, we presented a scalable, lane-wise highway traffic anomaly detection framework that operates exclusively on video-based data. By leveraging AI-driven vision techniques, our system extracts interpretable lane-specific traffic features, including vehicle count, occupancy, and truck percentage. We introduced a novel dataset derived from real-world highway surveillance, capturing fine-grained traffic dynamics across multiple time periods and anomaly types. The framework integrates deep learning, rule-based logic, and machine learning methods to detect both road-independent and road-dependent anomalies. Experimental results demonstrate the robustness and effectiveness of our approach in accurately identifying traffic anomalies, supporting its practical applicability in intelligent transportation systems.

\section*{Acknowledgment}

This work was supported by the Joint Transportation Research Program (JTRP), administered by the Indiana Department of Transportation and Purdue University, Grant SPR-4930. The authors would like to thank all
INDOT Study Advisory Committee members, for their guidance and advice throughout the project. The authors also would like to thank Chatgpt 4o for helping the grammar checking.

\bibliographystyle{IEEEtran}
\bibliography{ref}

\end{document}